\documentclass[10pt, conference, compsocconf, letterpaper]{IEEEtran}
\ifCLASSINFOpdf
  \usepackage[pdftex]{graphicx}
  
\else
\fi

\usepackage{amsmath}
\usepackage{algorithm}
\usepackage[noend]{algpseudocode}

\usepackage[utf8]{inputenc}
\usepackage[english]{babel}
 
\usepackage{amsthm}
 
\theoremstyle{definition}
\newtheorem{definition}{Definition}

\makeatletter
\def\ps@IEEEtitlepagestyle{
  \def\@oddfoot{\mycopyrightnotice}
  \def\@evenfoot{}
}

\def\mycopyrightnotice{
  {\footnotesize
  \begin{minipage}{\textwidth}
  \centering
  Copyright~\copyright~2019 IEEE.  Personal use of this material is permitted.  Permission from IEEE must be obtained for all other uses, in any current or future media, including reprinting/republishing this material for advertising or promotional purposes, creating new collective works, for resale or redistribution to servers or lists, or reuse of any copyrighted component of this work in other works.
  \end{minipage}
  }
}

\usepackage[font=footnotesize]{subfig}
\begin{document}
%
\title{
Long-term IaaS Provider Selection using Short-term Trial Experience}


\author{\IEEEauthorblockN{Sheik Mohammad Mostakim Fattah, Athman Bouguettaya, and Sajib Mistry }
\IEEEauthorblockA{School of Computer Science, University of Sydney, Australia\\
Email: \{sfat5243, athman.bouguettaya, sajib.mistry\}@sydney.edu.au} 
}

\maketitle

\begin{abstract}

We propose a novel approach to select privacy-sensitive IaaS providers for a long-term period. The proposed approach leverages a consumer’s short-term trial experiences for long-term selection. We design a novel equivalence partitioning based trial strategy to discover the temporal and unknown QoS performance variability of an IaaS provider. The consumer’s long-term workloads are partitioned into multiple Virtual Machines in the short-term trial. We propose a performance fingerprint matching approach to ascertain the confidence of the consumer’s trial experience. A trial experience transformation method is proposed to estimate the actual long-term performance of the provider. Experimental results with real-world datasets demonstrate the efficiency of the proposed approach.


\end{abstract}


%
\IEEEpeerreviewmaketitle

\section{Introduction}

Cloud computing is a key technology of choice for small to large organizations to establish and manage their IT infrastructures \cite{chaisiri2012optimization}. Cloud provides a faster, and cost-effective way to migrate in-house IT infrastructures. A large number of organizations manage their IT infrastructures in the cloud to achieve economy of sale. Large organizations such as governments, universities, banks subscribe to cloud services over a \textit{long-term} period (e.g., more than a year) \cite{ye2016long}. 

Infrastructure-as-a-Service (IaaS) is a primary service delivery model in the cloud. IaaS models typically offer computational resources such as CPU, memory, storage, and network bandwidth in the form of Virtual Machines (VMs). Amazon, Google, and Microsoft are examples of popular IaaS providers. The \textit{IaaS provider selection} for a long-term period is a topical research issue in cloud computing \cite{mistry2016qualitative}.

The performance of IaaS providers plays an important role in the selection of IaaS providers. The IaaS performance is often measured in terms of its Quality of Service (QoS) such as price, throughput, and availability. A consumer generally concerns two key aspects of the IaaS performance for the long-term selection. First, how the provider may perform under the consumer's \textit{long-term workloads}. The performance of IaaS providers usually varies depending on the workloads \cite{iosup2011performance}. Second, how the performance may vary over the long-term period for its workloads. Most IaaS providers are reluctant to reveal much information about their performance to protect themselves from their competitors. We define this unwillingness of revealing information as the \textit{privacy-sensitiveness} of IaaS providers. Privacy-sensitiveness is an intrinsic nature of IaaS providers that restrict them to divulge detailed and complete information about their services. The main reasons for such privacy-sensitiveness are market competition and business secrecy \cite{binnig2009weather}.  

Most existing studies mainly focus on short-term IaaS provider selection approaches \cite{mistry2018metaheuristic}. These approaches rely on IaaS advertisements for the selection process and are not applicable to select privacy-sensitive IaaS providers. IaaS advertisements typically contain \textit{incomplete} and \textit{convoluted} information to protect providers' \textit{business privacy}. For instance, Amazon AWS mentions only the availability of a service in its advertisements. Information about throughput, response time is not available in its advertisements. IaaS providers often advertise average or maximum performance information of their services. For instance, Amazon EC2 A2 instance advertise its network performance up to 10 Gbps. A consumer may not rely on such advertisements as actual performance is not guaranteed.  

Several studies introduce application and micro-benchmarks to predict the performance of IaaS providers according to consumer requirements \cite{scheuner2018estimating}. Application benchmarks are utilized to evaluate providers using different applications such as web applications, database applications. Micro-benchmarks reveal the performance of individual resources of VMs such as CPU, memory, and network bandwidths \cite{ostermann2009performance}. These approaches do not consider the \textit{long-term performance variability} of IaaS providers.

IaaS providers in the cloud market offer free trials for their services. For example, Microsoft Azure offers \$200 credit for 30 days for a limited number of services. Although IaaS providers do not explicitly share detailed information about their services, consumers may get a first-hand experience about IaaS providers using the free trial periods. To the best of our knowledge, existing studies do not consider the effective utilization of free trial periods for the long-term IaaS provider selection. \textit{We aim to utilize free trial periods to find out unknown QoS performance information of IaaS providers for the long-term IaaS provider selection.}

There are two main challenges of using trial periods for long-term selections. First, IaaS providers typically offer free trial periods for \textit{short-term} periods with limited flexibility. The consumer can not test its long-term workloads in such short trial periods.  An \textit{unplanned} utilization of such short-term trial periods may not properly reflect the actual performance of the provider. For example, if the workloads of a consumer have a long-tailed distribution, a one-month trial with a balanced request distribution may not divulge the true performance of long-tailed workloads. Second, the performance information found in the trial periods is applicable for a short-term period. The performance of public IaaS providers varies over time  due the \textit{dynamic} and \textit{chaotic} nature of the cloud environment \cite{leitner2016patterns}.  

The performance observed in trial periods primarily depends on the consumer's workloads and the provider's performance at that time. Both of these factors should be taken into account while performing the trial. \textit{We propose a novel trial strategy based on an equivalence partitioning method to capture the effect of the consumer's workloads on the provider's performance while considering the provider's temporal performance behaviour.}

We utilize the concept of \textit{performance fingerprint} for the long-term selection. The performance fingerprint of an IaaS provider represents an aggregated view of its temporal performance behavior. \textit{We assume the performance fingerprints of IaaS providers are known in this work}. \textit{We propose a fingerprint matching technique to ascertain the confidence of the consumer's trial experience for long-term selection.} If the trial experience of a consumer is consistent with a provider's performance fingerprint, we utilize the fingerprint to predict the provider's long-term performance for the consumer's long-term workloads. The trial experience may not entirely match the performance fingerprint as it represents an aggregated view of the provider's performance regardless of the consumer's workloads. The provider may provide an isolated trial environment where a consumer may not be able to observe its actual performance. \textit{We propose a trial experience transformation technique using the provider's performance fingerprint to estimate the actual performance of the provider for the consumer's workloads.} Our contributions in this work are as follows:

\begin{itemize}

\item An equivalence partitioning based trial strategy using a time series compression technique that maps the consumer's long-term workloads into multiple VMs in a short-term trial period to discover a privacy-sensitive IaaS provider's unknown QoS performance.  


\item A performance fingerprint matching technique to ascertain the confidence of the consumer's trial experience using the providers' performance fingerprints. 

\item A long-term performance discovery approach to select privacy-sensitive providers using time-series analysis. 

\end{itemize}

\section{Motivation Scenario}

Let us assume a university requires some general purpose VMs for one year where each VM has at least 2 vCPU and 4 GB memory.  The required number of VMs, resource requirements for each VM are considered as the functional requirements of the university. We assume the university has deterministic workloads, i.e., workloads are known for one year. The university represents the workloads in terms of the number of requested resources per day. The workloads may change over time depending on the number of students, holiday periods, and so on. The university defines minimum QoS requirements on throughput, response time, and availability of the VMs. The QoS requirements may also vary over time depending on seasonal demands.



Let us assume there are three IaaS providers Google, Amazon, and Microsoft who fulfil the university's functional requirements. No providers advertise their long-term performance on throughput, response time, and availability. We assume each provider offers a one-month free trial period to the university and allows the university to use three VMs. The university may run some representative benchmarks on three VMs for each day of the one-month trial period and monitor the performance of each provider to make the selection. It may lead to poor decision making as it does not consider the university's long-term workloads and the providers' temporal performance behaviours. The performance of a provider may fluctuate in the trial period. The university requires an effective trial strategy to understand the effect of different types of workloads on the provider's performance while considering the provider's temporal performance behaviour. 


We assume that the performance fingerprint of each provider is known to the university. The performance fingerprint provides the university with an aggregated view of a provider's temporal performance behaviour regardless of any specific type of workload distribution. Hence, the university requires to evaluate the performance of the providers using its workloads. If the provider's performance fingerprint and the trial experience exhibit similar temporal performance behaviour, the university may use the trial experience to evaluate the provider with high confidence. The university requires a fingerprint matching technique to evaluate its trial experience. We propose a set of tools in this paper that enables the university to leverage trial periods effectively to make an informed decision for the long-term period.

\section{The Proposed Framework}

We identify the following key challenges for the long-term selection using free trial periods:

\noindent \textit{(1) Restriction on Trial Periods}: IaaS providers assign different types of restricted condition on free trial periods:

\begin{itemize}
 \item Free trial periods are typically offered for a \textbf{short-term period}. Discovering long-term performances directly from short-term trials may not be possible. Amazon offers a one year trial period for some services. A consumer may not be able to wait such a long period to discover performances for the IaaS selection.


\item Most providers offer trial periods only for a limited number of services. For example, Amazon allows a user to give trial only t2 instances from EC2 VMs. The required types of VMs of a consumer may not be available for trial. In such a case, the consumer might be provided with similar yet different types of VMs for the trial. IaaS providers also restrict the number of available VMs for trial. 
\end{itemize}


\noindent \textit{(2) Temporal Performance Variability:} The performance discovered in the short-term trial periods may not always reflect the actual performance of the provider.  Almost all public IaaS providers typically use multi-tenant environments to provide services to their consumers. The effect of multi-tenancy on the performance may depend on several factors such as location, workloads on the provider, and QoS management strategy that vary with time. Multi-tenancy management policies are not revealed publicly due to the \textit{privacy-sensitiveness} of the providers. \textit{The measured performance in the trial in one month may be different in another month}. 


\noindent \textit{(3) Isolated Trial Environment:} IaaS providers may use an isolated environment for the trial users. In such a case, the trial consumers do not perceive the experience of a real cloud environment.\textit{ The consumers require a way to find whether they are treated differently than the existing consumers.}

\begin{figure}
    \centering
    \includegraphics[width=.48\textwidth]{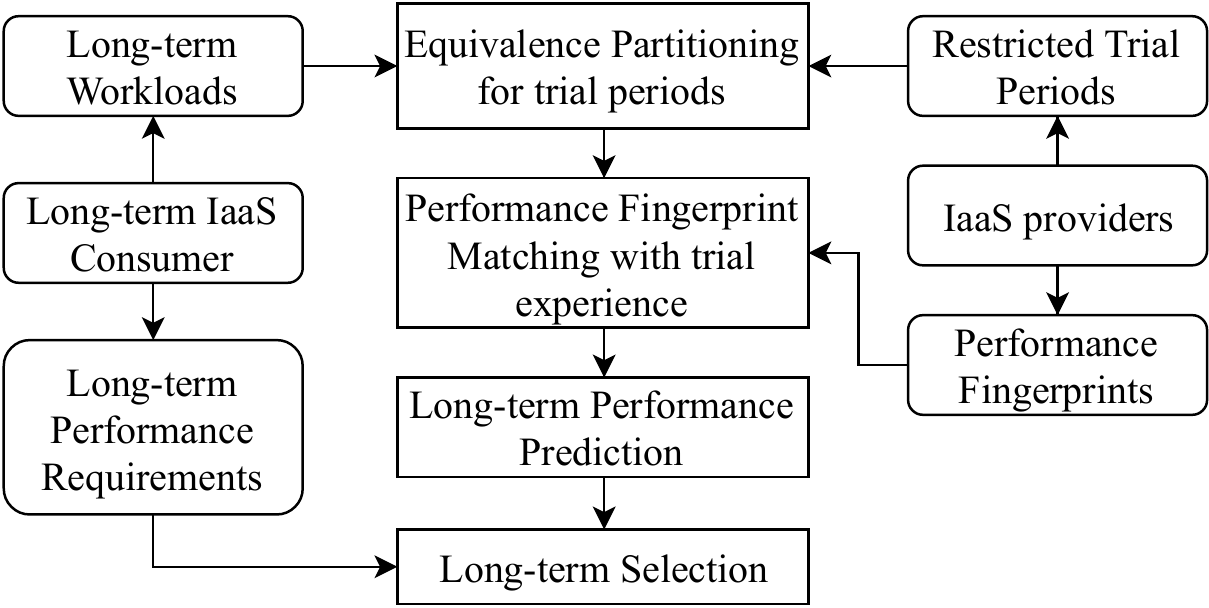}
    \caption{Long-term IaaS Provider Selection Framework}
    \label{fig:frame}
    \vspace{-.5cm}
\end{figure}

Fig. \ref{fig:frame} shows an IaaS provider selection framework that takes a consumer's long-term workload and the performance fingerprints of the providers to perform the selection. First, the proposed framework generates trial workloads using an equivalence partitioning method. Next, A performance fingerprint matching technique is applied to the trial experience to ascertain its confidence. The trial experience is then used for long-term performance prediction using the providers' fingerprints. Finally, the framework selects providers based on the consumer's long-term performance requirements. We discuss each of these steps in the following sections.

\section{An Equivalence Partitioning based Trial Strategy}



We define an equivalent partitioning based trial strategy where the consumer's workloads are tested in the trial period to discover a provider's performance while considering the provider's temporal performance fluctuation. For simplicity, we assume that the providers offer a fixed number of required VMs trial period using a \textit{continuous time-based model}. We consider the long-term workloads as time series data. We utilize time series compression techniques to capture the essential characteristics of the university's long-term workloads and map these workloads into the multiple VMs in the trial period.  

\subsection{Trial Workload Generation for Multiple VMs}

Let us assume the university's long-term workloads has $n$ number of workload data points i.e., $t_1, t_2, t_3,...,t_n$ over $T$ period. For instance, the university defines the workload as the average number of requested resources per day for one year. Each provider offers $v$ number of VMs for $T_r$ trial period. \textit{We assume that the performance fluctuation within  $d$ period is negligible.} Hence, workloads for a particular VM should remain same for every $d$ period over $T_r$ period to understand the effect of temporal performance behavior of a provider.  Each VM may run different types of workloads to understand the effect of provider's performance for different types of workloads. This method of partitioning the workloads is called \textit{equivalent partitioning.}

The university's long-term workloads of $n$ size need to be mapped with $v$ number of VMs on $T_r$ period. First, we partition the workloads into $n/v$ equal parts. Let us assume each part $w$ contains $m$ workload data points. Once we allocate the workloads for each VM, we need to compress the workload as $w$ may be still very large to run in $d$ period. The size of $w$ may be still too large to run in $d$ period. For example, if the university has one-year workloads and the number of VMs are 12, each part of the workload contains one month of workloads. Each VM should run one month of workloads on every day ($d=1$) of the trial period $T_r$. 

We need to compress each $w$ into $d$ period for each VM. Let us assume $d$ can be divided into $k$ data points. If $m \leq k$ then each workload of $w$ can be tested in the $d$ period. If $m > k$, then there are more workloads then compression is required. The compression may incur some loss of workload information. The loss depends on \textit{the size of $k$},\textit{ the shape of the workload time series}, and \textit{the compression method} \cite{burtini2013time}. 


We use a compression technique $M$ to extract the most important workloads from $w$ and fit into $d$ periods. $L()$ is the loss function that calculates the loss incurred during the compression using $M$ and $th$ is the maximum acceptable loss. During compression $L(M)\leq th$ condition must be held. $th$ is defined by the consumer. Let us assume $d$ can be divided into $k-1$ intervals. The method $M$ compresses $w$ to fit into $k-1$ intervals. Our target is to find an optimal value for $k$ where $L(M) \leq th$. We use Algorithm \ref{alg:workload} to generate workload for the $d$ period. The algorithm takes each $w$ from the university's long-term workload, the length of $d$, a compression technique $M$, a loss function $L()$, and minimum acceptable loss $th$. The algorithm produces trial workload $tw$ as output. First, it determines the initial value of $k$ by using a \textit{workload summarization} method. The workload summarization technique generates all different workloads for given workloads. The algorithm set the sampling rate $rate$ based on the size of the workload $m$ and the initial value of $k$. The algorithm then applies the compression method $M$ using $rate$ and calculates the amount of loss using $L()$. If it is less than an acceptable threshold, this process continues until the amount of loss $error$ is less than the maximum acceptable loss of $th$. Once $error \ge th$, the algorithm stops and returns the trial workload $tw$. 

\begin{algorithm}
    \caption{Generating Trial Workloads}\label{alg:workload}
    \begin{algorithmic}[1]
       
        \State \textbf{Input: }$w$, $d$, $M$, $L()$, $th$
        \State \textbf{Output: }$tw$ 
        \\
        
        $m \leftarrow size(w)$;\\
        $workloadSummary \leftarrow summary(w)$;\\
        $k \leftarrow size(workloadSummary)$;\\
        $rate \leftarrow ceil(m/k)$;\\
        $errorThresh \leftarrow th$;\\
        $error \leftarrow 0$;
        \While{$error<errorThresh$}
            \If{$rate==size(w)$}
            \State break;
            \EndIf
            \State $tw \leftarrow M(w,rate)$
            \State $error \leftarrow L(w)$
            \State $rate = rate+1$       
              
        \EndWhile   
        \\
        return $tw$;
    \end{algorithmic}
\end{algorithm}
\vspace{-5mm}

\subsection{Workload Compression Technique}

The university's long-term workloads are represented using time series. Several approaches exist to compress time series data and generated representative time series. We decide to use a commonly used time series compression technique called Piecewise Aggregate Approximation (PAA). The PAA method reduces the number of data points in a time series by taking average values in each interval ($I_i$). If $t_i$ represents a timestamp in the workload time series and $m$ is the total number of points in the time series, then the value of the time series in $I_j$ is calculated using the following equation: 

\begin{equation}
I_j = \frac{1}{x} \sum_{i=(j-1)*x+1}^{i*x}t_i  \text{ for } j = 1...\lceil m/x  \rceil
\label{eqn:compress}
\end{equation}
The size of the original time series can be reduced by any factor by changing the value of $x$. For a given $m$ data points in the workloads and $k$ segments in the trial period, we define the minimum $x=\lceil m/k \rceil$. The PAA method introduces some loss of information. For two given time series $w$ and $z$, the loss is defined by the following equation:


\begin{equation}
    L = \frac{1}{m} \sum_{i=1}^n |z_i-w_i|
    \label{eqn:loss}
\end{equation}

The original workload time series and the compressed workload time series has a different number of workload points. We decompress the compressed workload time series to compare with the original workload time series. We find $k$ workload points using equation \ref{eqn:compress} during the compression. We apply a decompression mechanism on the compressed workload of $k$ points to generate $n$ points. The decompression is performed by mapping each value of the compressed time series with $\lceil m/k \rceil$ interval into $m$ space. The rest $m-k$ points are generated by the linear extrapolation method. After decompression both original and compressed workload time series have the same number of points. Hence, we can apply equation \ref{eqn:loss} to compute the \textit{mean absolute error}.

\section{Performance Fingerprint Matching}

The performance information found in the trial periods is applicable for a short-term period. Many organizations such as CloudSpectator, CloudHarmony, and CloudStatus are devoted to monitor and analyze the performance of public IaaS cloud providers due to its growing importance. These organizations publish reports on the performance of IaaS providers using standard benchmarks. Each provider shows unique performance characteristics and exhibits different temporal performance behavior over the long-term periods. Each provider has its own unique temporal performance behavior that may depend on its provisioning policy, the number of consumers, and location.


We leverage the idea of fingerprinting to represent the temporal performance behavior of a provider. Fingerprinting techniques are well-known to identify and track a user on the Internet based on the impression left by the user \cite{takeda2012user}. Fingerprinting techniques are typically used to partially or fully identify a user on the Internet by tracking its activity and preferences without any active identification. We use the concept of \textit{performance fingerprint} of an IaaS provider to represents an aggregated view of the provider's long-term performance behavior. 

\theoremstyle{definition}
\begin{definition}{\textit{Performance Fingerprint}}
A performance fingerprint of an IaaS provider is the average performance a set of QoS parameters for each time interval over a fixed period that captures the provider's temporal performance behavior.
\end{definition}

We denote the performance fingerprint as $F = \{Q_1,Q_2,..Q_N\}$ where $N$ is the number of QoS parameters and $Q_i = {(P_n, t_n)| n = 1,2,3,..k}$. Here, $t_n$ denotes a timestamp of $T$ period where the average performance of $Q_i$ is $P_i$. The performance fingerprint of a provider may be known partially or completely. The partial fingerprint refers to a fingerprint that does not have information for all timestamps of a certain period.

\subsection{Performance Fingerprint Matching} 

We utilize the performance fingerprint of each provider to ascertain the confidence of the trial experience. If the trial experience is consistent with a provider's performance fingerprint, then the consumer may make the selection with confidence based on the trial experience.  

We assume the complete performance fingerprints of the providers is known for $T$ period. The trial is performed in the interval $(t_j,t_k) \in T$ i.e., $T_r=(t_j,t_k)$ where $j<k$ and $j,k \in T$ for a set of VM $v=\{v_1,v_2,...v_p\}$. The trial performance observed by the consumer for each VM is $Q_{vi} = \{q_1,q_2,..q_c\}$ where $c$ is the number of QoS parameters in the consumer requirements and $q_i=\{(p_n,t_n)|n=t_j,...t_k\}$ where $p_n$ is the performance of $q_i$ at the timestamp $t_n$. The first step of fingerprint matching is to aggregate the performance of each VM $v_i$ for each QoS parameter $q_i \in Q_{vi}$. The aggregated performance for each QoS parameter is computed by the following equation:

\begin{equation}
    q'_i = \text{sum} ( v_1 (q_i), v_2 (q_i),..., v_p(q_i))
    \label{eqn:qos}
\end{equation}
where $sum()$ represents the aggregate function, $v_j (q_i)$ represents the performance time series of QoS parameter $q_i$ in the VM $v_j$ in the trial period. The aggregated QoS performance of $q_i$ for all VM is $q'_i$. 

We denote the performance of the trial period for the consumer's aggregated workloads as $Q_{VM} = \{q'_1,q'_2,..q'_c\}$. Now, we need to perform fingerprint matching between $Q_{VM}$ and $F$ for the trial interval $(t_j,t_k)$. We use \textit{Pearson correlation coefficient} to compute the similarity between the trial experience and the performance fingerprint for each QoS parameters using the following equation:

\begin{equation}
    r_{q'_i,q_i} = \frac{\sum_{t=j}^k (p'_t - \Bar{p'}) (p_t - \Bar{p})}{\sqrt{\sum_{t=j}^k (p'_t - \Bar{p'})^2} \sqrt{\sum_{t=j}^k (p_t - \Bar{p})^2}}
    \label{eqn:pearson}
\end{equation}

where $p'_t$ the value of observed performance and $p_t$ is the value of the performance fingerprint at the time $t$ of the trial period for a QoS parameter.  The mean correlation coefficient for all QoS parameters is computed as follows:

\begin{equation}
    R_{Q_{VM},F} = \frac{1}{c}\sum_{i=1}^c r_{q'_i,q_i} 
\end{equation}

The correlation coefficient measures the similarity of two time series in terms of their trends, i.e., how much the trial experience is affected by the performance fingerprint. It does not consider the actual distance from the fingerprint. We define the confidence of the trial by considering both trend and distance of the trial experience with the fingerprint using the following equation:

\begin{equation}
    \text{Confidence} = (R_{Q_{VM},F},\text{MNRMSE}(Q_{VM},F))
    \label{eqn:confidence}
\end{equation}

where MNRMSE is the \textit{mean normalized root mean squared error} between the trial experience and the performance fingerprint. First, we compute the NRMSE for each QoS parameter in the trial period. The MNRMSE is computed by taking average NMRSE of all QoS parameter. The consumer defines a minimum threshold $R_t$ and $E_t$ for $R_{Q_{VM},F}$ and $\text{MNRMSE}(Q_{VM},F)$ respectively. If the confidence of the trial experience is below the thresholds, we consider it as a partial fingerprint matching. 






\subsection{Trial Experience Transformation for Partial Matching}

We transform the trial experience for partial fingerprint matching to estimate an approximate performance behavior of the provider for the consumer's workloads. The trial experience may have less correlation or higher distance with the performance fingerprint of a provider. We need to transform the trial experience in a way that it reduces the distance or increase the correlation between trial experience and the performance fingerprint. In both cases, the confidence of the trial may increase. 

The partial fingerprint matching indicates that the actual performance of the provider may be different from the trial experience for the consumer's workloads. We use the following equation to transform the trial experience to estimate the actual performance:

\begin{equation}
    Q_{VM}^T = Q_{VM}+\frac{1}{2}(F - Q_{VM})
    \label{eqn:transform}
\end{equation}

where $Q_{VM}^T$ is the transformed trial experience, $F$ is the performance fingerprint in the trial interval, and $Q_{VM}$ is the trial experience. Equation \ref{eqn:transform} transforms the trial experience for each aggregated QoS parameters by reducing the distance from the fingerprint by half. The intuitive idea behind this transformation is two-fold. First, if the provider offers an isolated trial environment, the real experience may be closer to the fingerprint rather than the trial experience. Second, the performance fingerprint does not contain information for the consumer's workload distribution. The transformation increases the confidence of the trial experience.

\section{Long-term IaaS Provider Selection}

We discuss the long-term selection process using the trial experience and the provider's performance fingerprint in this section. First, we estimate providers' long-term performance for the consumer's workloads using the trial experience and the performance fingerprints. Next, we rank the providers based on their performance and the consumer's long-term performance requirements.


\subsection{Long-term Performance Discovery}

The university's important workloads are tested in the trial periods and the required QoS parameters are monitored. Let us assume the trial workloads $TW = \{tw_1,tw_2,....,tw_k\}$ are tested in $v$ number of VMs. Each type of workload is  monitored in the trial period $T_r$ for each $d$ intervals where the performance fluctuation in $d$ is negligible. The QoS performance for each workload is denoted by $Q_{tw_i}=\{q_1,q_2,..,q_c\}$ where $q_i=\{(p_n,t_n)|n=1,d,...,Tr\}$ and $p_n$ is the performance observed at the timestamp $t_n$. The consumer's long-term workloads is denoted by $LW = \{W_1,W_2,....,W_T\}$. We need to find the performance for each $W_i$ which is denoted by $Q_{W_i}$. The trial performance $Q_{tw_i}$ and the performance fingerprint $F$ are used to generate $Q_{W_i}$. We use the following steps to compute $Q_{W_i}$:

 \begin{enumerate}
     \item For each $W_i \in LW$, find the closest $w_i \in TW$. 
     \item For each $q'_i \in Q_{W_i}$, find $q_i \in Q_{tw_i}$.
     \item Let $t'_i$ is the timestamp of $W_i$. $tw_i$ has $Tr/d$ number of observations. We select a timestamp $t_i$ for $tw_i$ where $(t_i-t_1)=(t'_i \; \text{mod} \; (T/Tr))$ where $T$ is the total time and $Tr$ is the trial period. For example, if $Tr= 30$, $T = 360$, and $t'_i = 35$ then $t_i=5$.  
     
    \item We compute the relative weight $r_w$ of the fingerprint at $t_i$ for $q'_i$ using the following equation:
     \begin{equation}
         r_w = \frac{P_{t'_i}}{P_{t_i}}
     \end{equation}
     where $P_{t'_i}$ and $P_{t_i}$ is the performance of the fingerprint at timestamp $t'_i$ and $t_i$ respectively.
     \item The performance of $q'_i$ at $t'_i$ is computed as follows:
    \begin{equation}
         p'_{t'_i} = r_w * p_{t_i}
     \end{equation}
     where $p'_{t'_i}$ and $p_{t_i}$ is the performance $q'_i$ at timestamp $t'_i$ and $t_i$ respectively.
\end{enumerate}

We compute the relative weight of the performance fingerprint between the timestamp of real workload and the trial workload. The relative weight is applied to the trial performance of the particular QoS value to compute the performance of the real workload. We perform the above steps for each $q_i \in Q_{W_i}$ to generate the long-term performance for each provider.

\subsection{IaaS Provider Selection}

We compute the distance between the estimated performance of a provider and the consumer's long-term performance requirements. The rank of each provider is computed based on their distance from the consumer's long-term requirements. We use normalized root mean square distance to compute the distance for each QoS parameter using the following equation: 

\begin{equation}
    d(q_c,q_p) = \sqrt[]{\frac{1}{n}{\sum_{q \in q_c, q' \in q_p,{t=1..n}}  (q_t-q'_t)^2}} 
\label{eqn:distance}
\end{equation}

where $q_c$ and $q_p$ are the time series of the consumer's long-term requirements and the provider's estimated long-term performance for a particular QoS parameter respectively. The total distance for all QoS parameter is computed by the following:

\begin{equation}
    D(Q_c,Q_p) = \sum_{i=1}^c di(q_{ci},q_{pi})
\end{equation}

where $Q_c$ and $Q_p$ are the consumer's requirements and the provider's estimated performance respectively.

\section{Experiments and Results}

A set of experiments is conducted to evaluate the proposed approach. First, we show that the proposed trial strategy can predict a provider's long-term performance using its performance fingerprints. Next, we evaluate the effectiveness of the trial experience transformation technique considering partial fingerprint matching. Finally, we rank IaaS providers based on long-term performance prediction.

\begin{figure*}[t!]
 \centerline{
  \subfloat[]{\includegraphics[width=.5\textwidth]{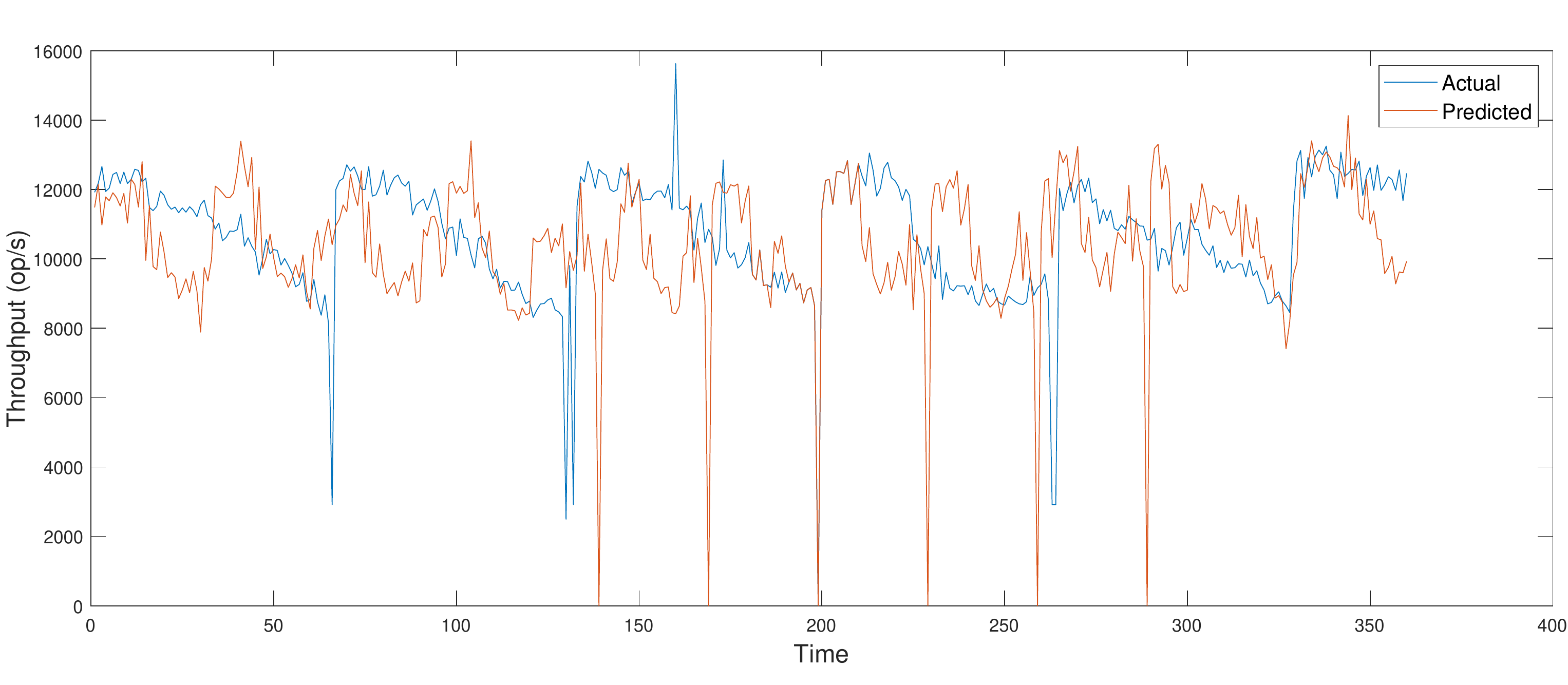} }
  \hfill
  \subfloat[]{\includegraphics[width=.5\textwidth]{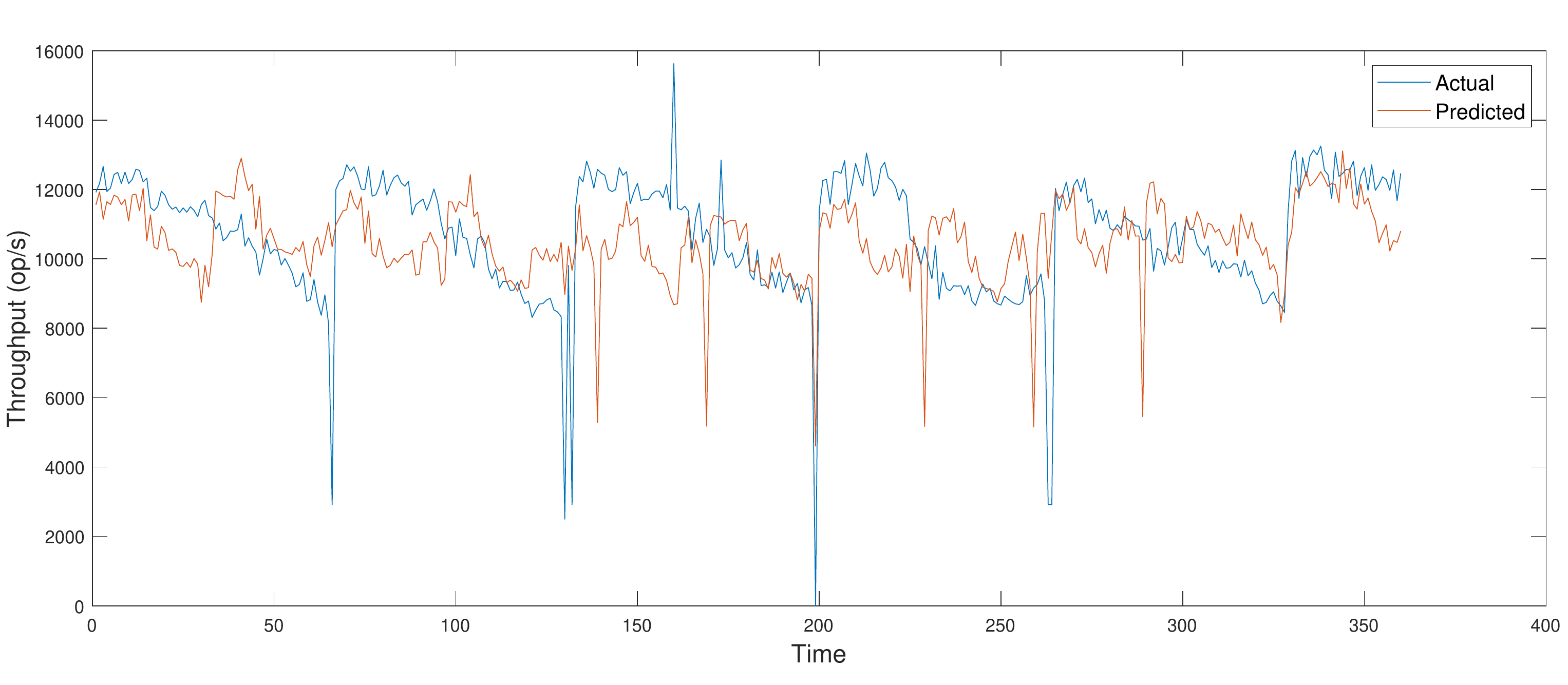}}
  }
   \centerline{
\subfloat[]{\includegraphics[width=.5\textwidth]{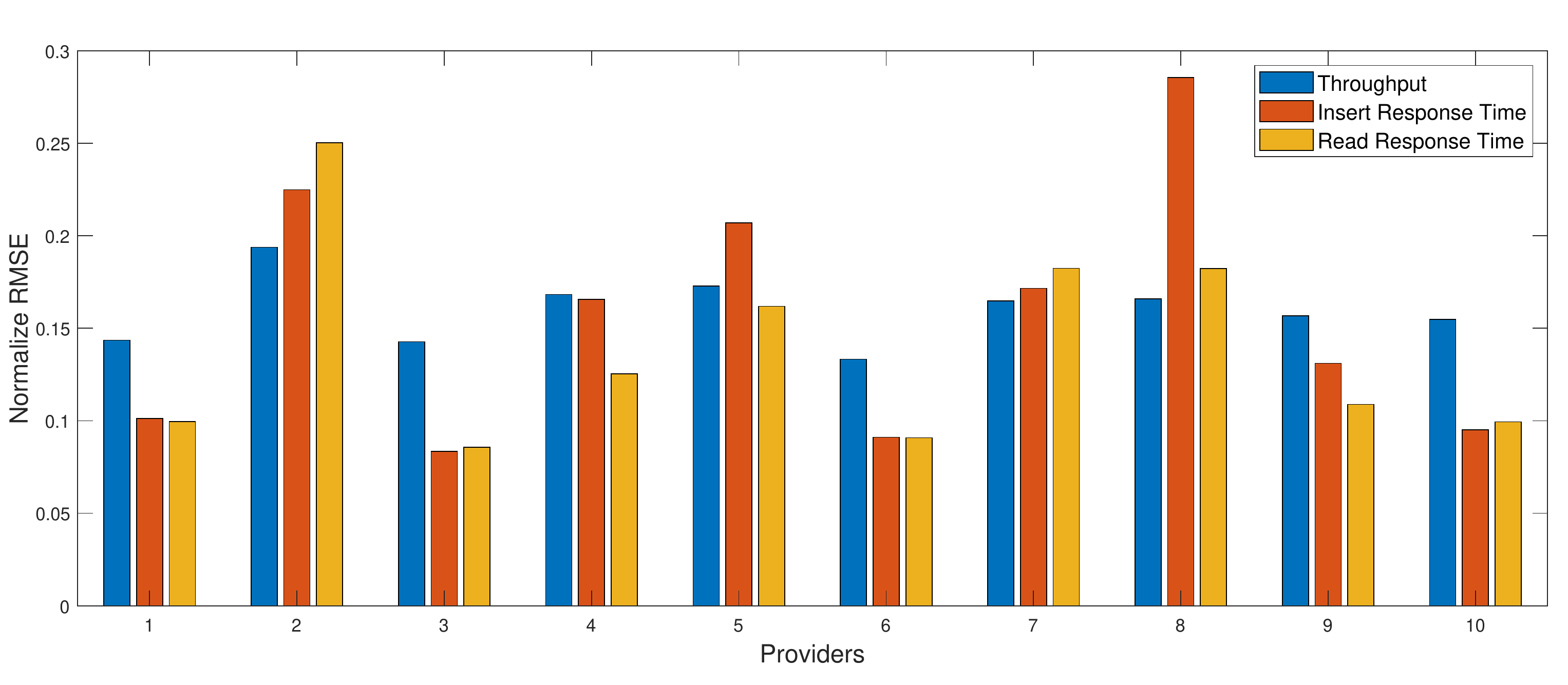}}
  \hfill
   \subfloat[]{\includegraphics[width=.5\textwidth]{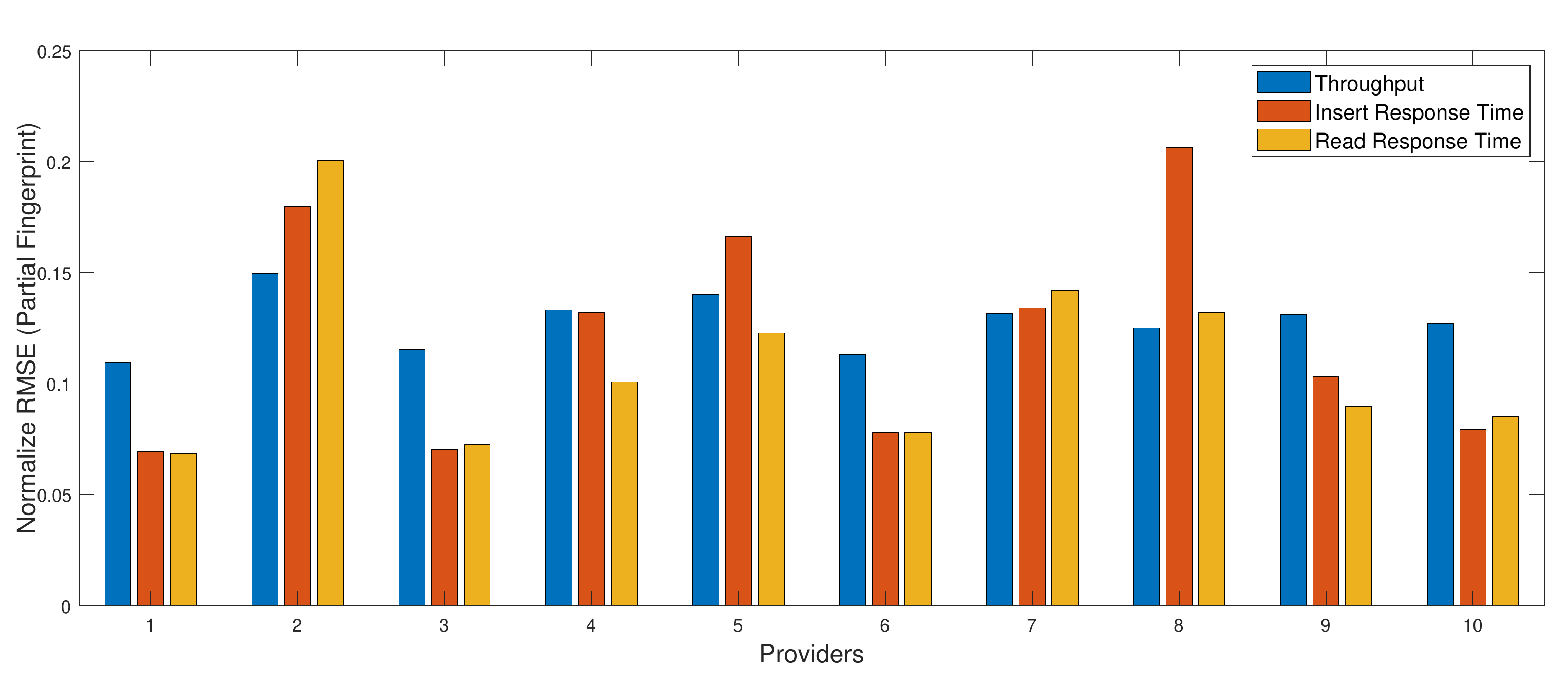}}
  }
\caption{Long-term performance prediction (a) Throughput (b) Throughput with partial fingerprint (c) Normalize RMSE prediction accuracy (d) Normalize RMSE prediction accuracy with partial fingerprint  }
\label{fig:prediction}
\vspace{-.5cm}
\end{figure*}

\subsection{Experiment Setup}
Finding real-world cloud traces for a long-term period is challenging. We generate the CPU workloads for the consumers from publicly available Eucalyptus cloud traces. It contains data of 6 clusters which cover continuous multi-month time frames \cite{nurmi2009eucalyptus}. We select one trace to generate CPU workloads for ten consumers. The QoS performance data is collected from SPEC Cloud IaaS 2016 results \cite{baset2017spec}.  We generate QoS performances of each provider for each consumer's workloads by random replication method. Data of one month are mapped into 12-month data points where each data point is considered as an average of a single day measurement. The performance fingerprint of each provider is generated by taking the average of the observed performances of all consumers. First, we select a consumer from ten consumers as new consumer. The trial data of the selected consumer is generated using the proposed approach for 12 virtual machines and 30 days.  First, we find the closest matched workload from other nine consumers to generate the performance data for the trial for each workload. The performance of the corresponding workload is considered as the trial performance of a new consumer. This approach ensures that the trial experience is affected by a provider's performance behavior. 


\subsection{Accuracy of the Performance Prediction}

Fig. \ref{fig:prediction} shows the results of a long-term performance prediction for a provider using its performance fingerprint. The performance fingerprints represent an aggregated view regardless of a consumer's workload. Hence, We can not predict the actual performance using only the provider's performance fingerprint. Fig. \ref{fig:prediction}(a) shows that the performance prediction without considering the trial experience transformation for the throughput of a provider. The performance prediction considering the partial fingerprint matching is shown in Fig. \ref{fig:prediction}(b). We use the confidence threshold $(0.5,1)$ for the similarity and distance respectively. Once we apply the transformation, the confidence of the trial increases significantly. The prediction accuracy also improves when partial fingerprint matching is considered. Fig. \ref{fig:prediction} depicts the long-term performance prediction for ten IaaS providers. Fig. \ref{fig:prediction}(c) shows the performance prediction without considering the partial fingerprint matching. Fig. \ref{fig:prediction}(d) shows the performance prediction considering the partial fingerprint matching. The prediction accuracy is higher i.e., lower \textit{Normalized RMSE distance} in Fig. \ref{fig:prediction}(c) than Fig. \ref{fig:prediction}(d) which proves that the performance prediction accuracy increases with the trial experience transformation technique.


\subsection{Accuracy of the Long-term Selection}
\begin{figure*}[t!]
 \centerline{
\subfloat[]{\includegraphics[width=.5\textwidth]{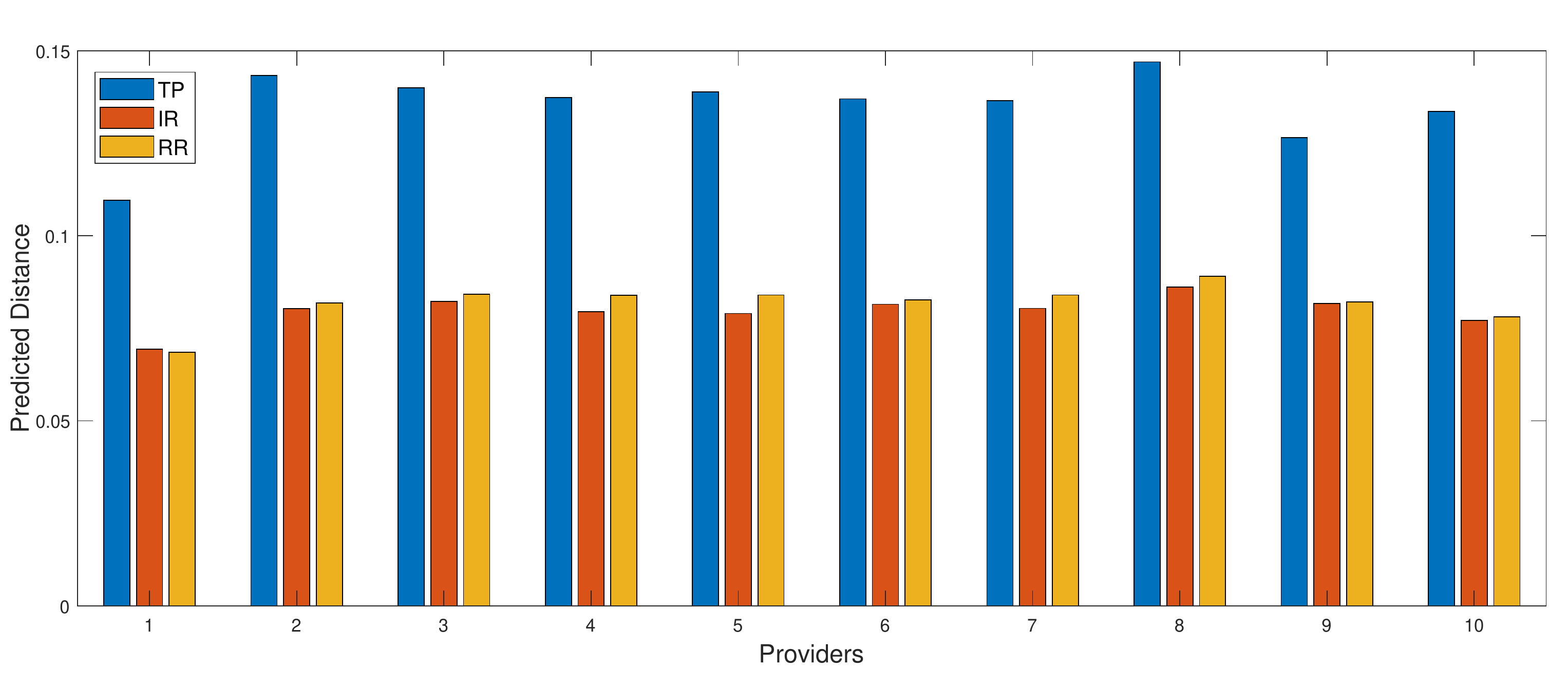} }
  \hfill
  \subfloat[]{\includegraphics[width=.5\textwidth]{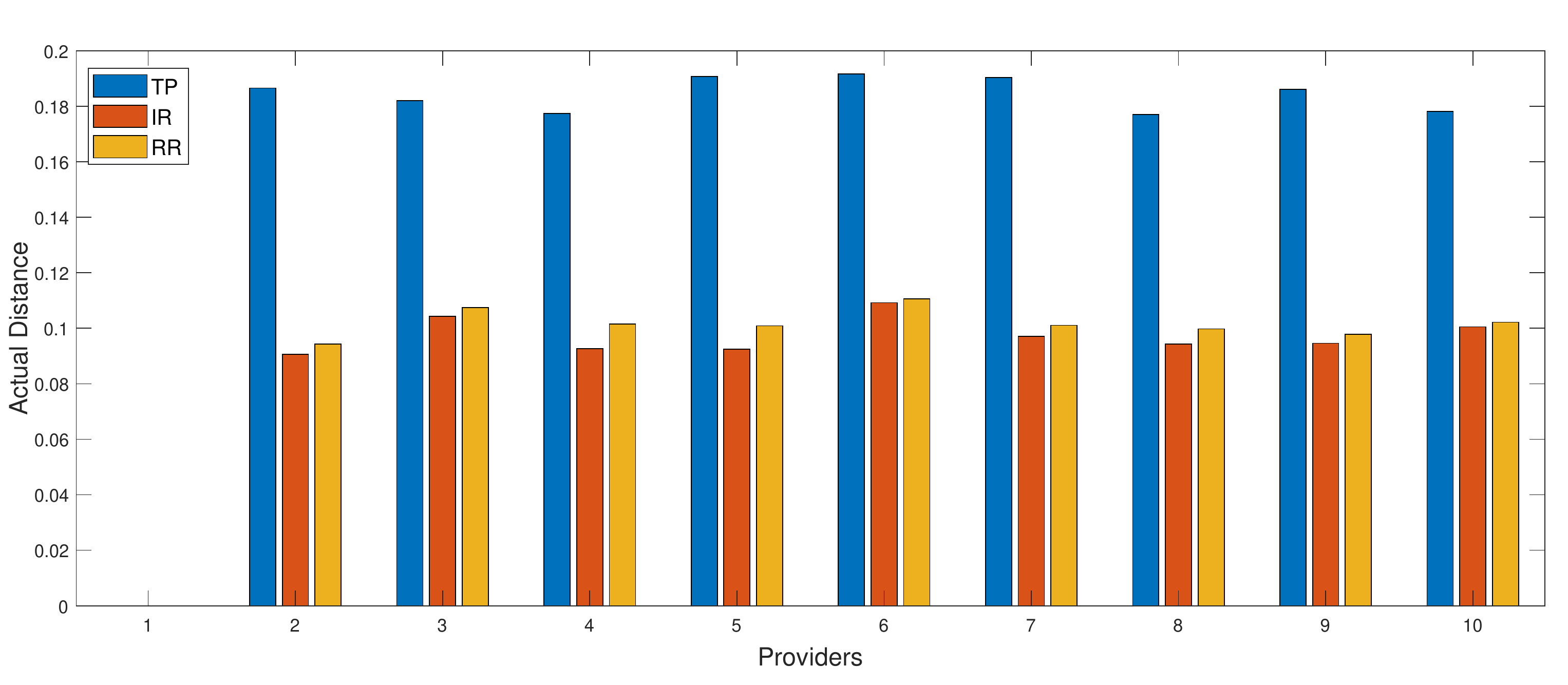} }
  }
\caption{Normalize RMSE Distance between provider and consumer (a) Predicted distance (b) Actual distance }
\label{fig:bar}
\vspace{-.5cm}
\end{figure*}

We use the \textit{Normalize RMSE distance} between a provider's performance and a consumer's requirements to rank each provider. The distance between the consumer's requirements and each provider's predicted performance is shown in Fig. \ref{fig:bar}(a). The figure shows that the provider 1 has the minimum distance from the consumer requirements for throughput, insert and read response time. Fig. \ref{fig:bar}(a) shows the distance between the providers' actual performance and the consumer's requirement. As we generated the consumer's requirement from provider 1's actual performance, provider 1 has zero distance from the consumer's requirement. Therefore, the proposed approach successfully select the optimal provider for the long-term period. 

\section{Related Work}

Several studies discover QoS performance of IaaS providers by deploying VMs in the cloud.  An extensive study on the performance variance of Amazon EC2 is provided in \cite{schad2010runtime}. It addresses that the performance unpredictability in the cloud is a significant issue for many users and often considered as a key obstacle in the cloud adaption. The study finds that Amazon EC2 shows high variance in its performance. The performance of clouds for scientific computing is analyzed using micro-benchmarks and kernels on Amazon EC2 in \cite{ostermann2009performance,iosup2011performance}. The proposed study observes that tested clouds are not suitable for scientific computing due to performance variance and low reliability.  Most studies conduct experiments to measure short-term performance \cite{bouguettaya2010end}. Existing performance monitoring and testing approaches do not consider the long-term selection. 

Fingerprinting is a well-known approach where a small portion of data is used to identify a data source uniquely. Fingerprinting is used in many computing domains such as public key management, digital video and audio copyright, digital forensic, and user tracking. A number of studies focus on passive fingerprinting technique to track users from their interaction in the browser without using cookies \cite{takeda2012user}.  These approaches focus on users from a provider's perspective. We gain insight from these approaches that temporal performance behaviour of IaaS provider is identifiable. We introduce the performance fingerprinting of IaaS providers in this work from a consumer perspective to capture their temporal performance variability.  



\section{Conclusion}

We propose a novel approach to select privacy-sensitive IaaS providers using their performance fingerprints. The proposed approach utilizes free trial periods to evaluate a provider's long-term performance. A consumer may choose a provider based on its trial experience. A novel trial strategy using equivalence partitioning method is proposed to estimate a provider's performance for different types of workloads while considering the provider's performance variability. The trial experience is incorporated with the provider's performance fingerprint to predict long-term performance. A performance fingerprint matching technique is proposed to ascertain the confidence of the consumer's trial experience. A trial experience transformation method is proposed to improve the confidence of the consumer's trial experience. The results of experiments show that our proposed approach helps a consumer to make an informed decision to select a privacy-sensitive IaaS provider for the long-term period. A key limitation is that we consider only a limited number of real-world IaaS providers. We aim to study the performance of a large number of IaaS providers to improve our proposed approach. 

\section{Acknowledgement}
This research was partly made possible by NPRP 9-224-1-049 grant from the Qatar National Research Fund (a member of The Qatar Foundation) and DP160103595 and LE180100158 grants from Australian Research Council. The statements made herein are solely the responsibility of the authors.


\bibliographystyle{IEEEtran}
\bibliography{IEEEabrv,reference}

\begin{thebibliography}{10}
\providecommand{\url}[1]{#1}
\csname url@samestyle\endcsname
\providecommand{\newblock}{\relax}
\providecommand{\bibinfo}[2]{#2}
\providecommand{\BIBentrySTDinterwordspacing}{\spaceskip=0pt\relax}
\providecommand{\BIBentryALTinterwordstretchfactor}{4}
\providecommand{\BIBentryALTinterwordspacing}{\spaceskip=\fontdimen2\font plus
\BIBentryALTinterwordstretchfactor\fontdimen3\font minus
  \fontdimen4\font\relax}
\providecommand{\BIBforeignlanguage}[2]{{%
\expandafter\ifx\csname l@#1\endcsname\relax
\typeout{** WARNING: IEEEtran.bst: No hyphenation pattern has been}%
\typeout{** loaded for the language `#1'. Using the pattern for}%
\typeout{** the default language instead.}%
\else
\language=\csname l@#1\endcsname
\fi
#2}}
\providecommand{\BIBdecl}{\relax}
\BIBdecl

\bibitem{chaisiri2012optimization}
S.~Chaisiri, B.-S. Lee, and D.~Niyato, ``Optimization of resource provisioning
  cost in cloud computing,'' \emph{IEEE TSC}, vol.~5, no.~2, pp. 164--177,
  2012.

\bibitem{ye2016long}
Z.~Ye, S.~Mistry, A.~Bouguettaya, and H.~Dong, ``Long-term qos-aware cloud
  service composition using multivariate time series analysis,'' \emph{IEEE
  TSC}, vol.~9, no.~3, pp. 382--393, 2016.

\bibitem{mistry2016qualitative}
S.~Mistry, A.~Bouguettaya, H.~Dong, and A.~Erradi, ``Qualitative economic model
  for long-term iaas composition,'' in \emph{ICSOC}.\hskip 1em plus 0.5em minus
  0.4em\relax Springer, 2016, pp. 317--332.

\bibitem{iosup2011performance}
A.~Iosup, N.~Yigitbasi, and D.~Epema, ``On the performance variability of
  production cloud services,'' in \emph{CCGrid}.\hskip 1em plus 0.5em minus
  0.4em\relax IEEE, 2011, pp. 104--113.

\bibitem{binnig2009weather}
C.~Binnig, D.~Kossmann, T.~Kraska, and S.~Loesing, ``How is the weather
  tomorrow?: towards a benchmark for the cloud,'' in \emph{Proceedings of the
  Second International Workshop on Testing Database Systems}.\hskip 1em plus
  0.5em minus 0.4em\relax ACM, 2009, p.~9.

\bibitem{mistry2018metaheuristic}
S.~Mistry, A.~Bouguettaya, H.~Dong, and A.~K. Qin, ``Metaheuristic optimization
  for long-term iaas service composition,'' \emph{IEEE Transactions on Services
  Computing}, vol.~11, no.~1, pp. 131--143, 2018.

\bibitem{scheuner2018estimating}
J.~Scheuner and P.~Leitner, ``Estimating cloud application performance based on
  micro-benchmark profiling,'' in \emph{CLOUD}.\hskip 1em plus 0.5em minus
  0.4em\relax IEEE, 2018, pp. 90--97.

\bibitem{ostermann2009performance}
S.~Ostermann, A.~Iosup, N.~Yigitbasi, R.~Prodan, T.~Fahringer, and D.~Epema,
  ``A performance analysis of ec2 cloud computing services for scientific
  computing,'' in \emph{ICCC}.\hskip 1em plus 0.5em minus 0.4em\relax Springer,
  2009, pp. 115--131.

\bibitem{leitner2016patterns}
P.~Leitner and J.~Cito, ``Patterns in the chaos—a study of performance
  variation and predictability in public iaas clouds,'' \emph{ACM TOIT},
  vol.~16, no.~3, p.~15, 2016.

\bibitem{burtini2013time}
G.~Burtini, S.~Fazackerley, and R.~Lawrence, ``Time series compression for
  adaptive chart generation,'' in \emph{CCECE}.\hskip 1em plus 0.5em minus
  0.4em\relax IEEE, 2013, pp. 1--6.

\bibitem{takeda2012user}
K.~Takeda, ``User identification and tracking with online device fingerprints
  fusion,'' in \emph{ICCST}.\hskip 1em plus 0.5em minus 0.4em\relax IEEE, 2012,
  pp. 163--167.

\bibitem{nurmi2009eucalyptus}
D.~Nurmi, R.~Wolski, C.~Grzegorczyk, G.~Obertelli, S.~Soman, L.~Youseff, and
  D.~Zagorodnov, ``The eucalyptus open-source cloud-computing system,'' in
  \emph{CCGRID}.\hskip 1em plus 0.5em minus 0.4em\relax IEEE, 2009, pp.
  124--131.

\bibitem{baset2017spec}
S.~Baset, M.~Silva, and N.~Wakou, ``Spec cloud iaas 2016 benchmark,'' in
  \emph{Proceedings of the 8th ACM/SPEC on International Conference on
  Performance Engineering}.\hskip 1em plus 0.5em minus 0.4em\relax ACM, 2017,
  pp. 423--423.

\bibitem{schad2010runtime}
J.~Schad, J.~Dittrich, and J.-A. Quian{\'e}-Ruiz, ``Runtime measurements in the
  cloud: observing, analyzing, and reducing variance,'' \emph{Proceedings of
  the VLDB Endowment}, vol.~3, no. 1-2, pp. 460--471, 2010.

\bibitem{bouguettaya2010end}
A.~Bouguettaya, S.~Nepal, W.~Sherchan, X.~Zhou, J.~Wu, S.~Chen, D.~Liu, L.~Li,
  H.~Wang, and X.~Liu, ``End-to-end service support for mashups,'' \emph{IEEE
  TSC}, vol.~3, no.~3, pp. 250--263, 2010.

\end{thebibliography}
%



\end{document}